\documentclass{article}
\usepackage{PRIMEarxiv}

\usepackage{cite}
\usepackage{amsmath,amssymb,amsfonts}
\usepackage{algorithmic}
\usepackage{graphicx}
\usepackage{textcomp}
\def\BibTeX{{\rm B\kern-.05em{\sc i\kern-.025em b}\kern-.08em
    T\kern-.1667em\lower.7ex\hbox{E}\kern-.125emX}}

\usepackage{xcolor, soul, tikz, svg, subcaption, enumitem, adjustbox, float}
\usepackage{array, caption, makecell, multirow, tabularx, tcolorbox, verbatim}

\usepackage[utf8]{inputenc} 
\usepackage[T1]{fontenc}    
\usepackage{hyperref}       
\usepackage{url}            
\usepackage{booktabs}       
\usepackage{amsfonts}       
\usepackage{nicefrac}       
\usepackage{microtype}      
\usepackage{lipsum}
\usepackage{fancyhdr}       
\usepackage{graphicx}       
\graphicspath{{media/}}     

\definecolor{lime}{HTML}{A6CE39}
\DeclareRobustCommand{\orcidicon}{
	\begin{tikzpicture}
	\draw[lime, fill=lime] (0,0) 
	circle [radius=0.16] 
	node[white] {{\fontfamily{qag}\selectfont \tiny ID}};
	\draw[white, fill=white] (-0.0625,0.095) 
	circle [radius=0.007];
	\end{tikzpicture}
	\hspace{-2mm}
}
\foreach \x in {A, ..., Z}{\expandafter\xdef\csname orcid\x\endcsname{\noexpand\href{https://orcid.org/\csname orcidauthor\x\endcsname}
			{\noexpand\orcidicon}}
}

\title{PersoPilot: An Adaptive AI-Copilot for Transparent Contextualized Persona Classification and Personalized Response Generation \\
}

\author{
    Saleh Afzoon $^{1}$ \orcidA{}, 
    Amin Beheshti $^{1}$\orcidB{},
    Usman Naseem $^{1}$ \orcidC{}\\
    $^{1}$School of Computing, Macquarie University, Sydney, Australia \\
    \small \texttt{saleh.afzoon@hdr.mq.edu.au},\small \texttt{\{amin.beheshti, usman.naseem\}@mq.edu.au}
}

\begin{document}
\maketitle

\begin{abstract}
Understanding and classifying user personas is critical for delivering effective personalization. While persona information offers valuable insights, its full potential is realized only when contextualized, linking user characteristics with situational context to enable more precise and meaningful service provision. Existing systems often treat persona and context as separate inputs, limiting their ability to generate nuanced, adaptive interactions. To address this gap, we present PersoPilot, an agentic AI-Copilot that integrates persona understanding with contextual analysis to support both end users and analysts. End users interact through a transparent, explainable chat interface, where they can express preferences in natural language, request recommendations, and receive information tailored to their immediate task. On the analyst side, PersoPilot delivers a transparent, reasoning-powered labeling assistant, integrated with an active learning–driven classification process that adapts over time with new labeled data. This feedback loop enables targeted service recommendations and adaptive personalization, bridging the gap between raw persona data and actionable, context-aware insights. As an adaptable framework, PersoPilot is applicable to a broad range of service personalization scenarios.
\end{abstract}


\section{Introduction}

Effective personalization in human-AI interaction requires not only an understanding of user traits but also the evolving context in which interactions occur. Traditional persona modeling often treats users as static profiles, neglecting dynamic goals, situational intent, and task-driven behavior. This limits the adaptability of AI assistants and reduces their effectiveness in delivering nuanced, context-aware interactions. We present PersoPilot, an adaptive AI copilot designed to integrate persona classification with real-time contextual reasoning and personalized response generation for both end users and analysts.

The importance of context in user modeling has already been established. For example, kBot \cite{kadariya2019kbot} leveraged domain knowledge and sensor data to improve personalized asthma interventions, while E-DHAP \cite{ma2021one} utilized implicit persona modeling from dialogue history to simulate agent-like behaviors. However, analyst-level interpretability and control are not central design goals in these systems, which primarily emphasize end-user personalization. More recent approaches have introduced hybrid modeling and explainable AI techniques to address these limitations. ExBigBang \cite{afzoon2025exbigbang}, for instance, combines textual and tabular data to support context-rich persona classification with interpretable outputs. CoBERT \cite{mpia2023cobert} further integrates structured context with deep pre-trained models to improve recommendation transparency and user understanding. E-ReDial \cite{guo2023towards} enhances user trust by generating natural language justifications for recommendations, though it remains focused on static domains without analyst-in-the-loop learning or dynamic persona updates.

PersoPilot extends these directions by introducing a dual-mode, context-aware framework that interleaves user interaction and analyst oversight. It treats persona development as a dynamic, task-sensitive process, tightly coupled with dialogue history and user intent. Rather than relying on static or exhaustive persona graphs, PersoPilot applies contextual filtering to retrieve only the most relevant traits and topics, streamlining knowledge grounding and enabling real-time, reasoning-augmented personalization. Through its labeling assistant and active learning loop, the system supports scalable annotation workflows while maintaining transparency and adaptability across both user-facing and analyst-driven interactions. As a generalizable framework, PersoPilot can be applied to any service personalization task where user information is available, broadening its utility across domains.

Our approach is informed by insights from PersoBench \cite{afzoon2024persobench}, which demonstrates that incorporating task-specific context significantly enhances personalization, and by few-shot prompt-based techniques \cite{afzoon2025modeling}, which we adapt for both user interaction and annotation workflows. By unifying contextual reasoning, explainability, and dynamic persona modeling, PersoPilot moves beyond traditional chatbots toward agentic copilots that support both end users and analysts through transparent, continuously adaptive personalization. A public implementation of the complete system, including the backend, web interface, and a demonstration video, is available at: \url{https://github.com/salehafzoon/PersoPilot}.

\section{System Architecture}

\begin{figure*}[!h]
\centering
  \includegraphics[width=0.77\textwidth]{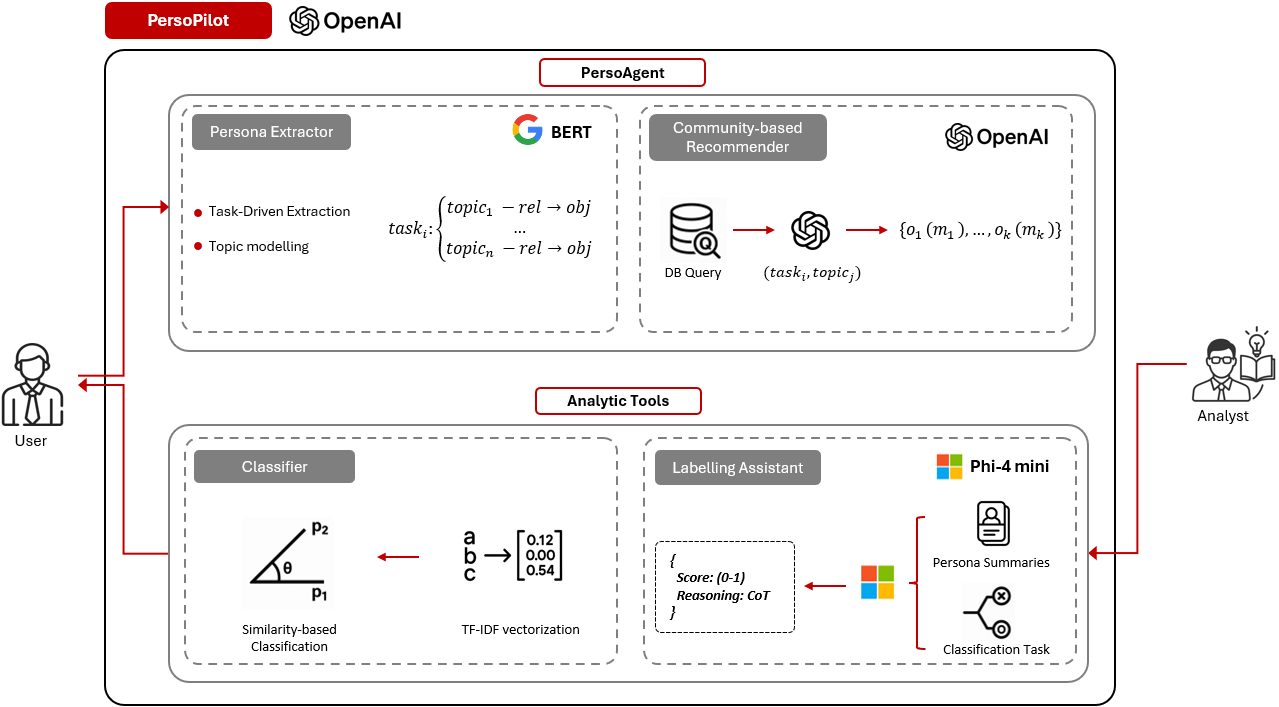}
  \caption{Architectural overview of the PersoPilot framework.}
  \label{fig:persopilot-overview}
\end{figure*}


Fig.~\ref{fig:persopilot-overview} provides an overview of the PersoPilot framework architecture, which is organized around two primary modules: \textit{PersoAgent}, an AI-driven agent for real-time user interaction and personalized recommendation, and \textit{Analytic Tools}, a backend suite supporting analyst-driven labeling, classification, and model refinement. The system functions as a dynamic feedback loop connecting user dialogue, contextual persona extraction, intelligent content filtering, and adaptive annotation, enabling continuous personalization through both user-facing and analyst-in-the-loop interactions.

\textbf{PersoAgent.} This module is implemented as an autonomous AI agent using the LangChain\footnote{\url{https://www.langchain.com/}} framework. The agent is powered by an OpenAI\footnote{\url{https://platform.openai.com}} model and integrates two internal tools: (1) a \textit{Persona Extractor} and (2) a \textit{Community-based Recommender}. The Persona Extractor is trained using the ConvAI2 dataset \cite{zhang2018personalizing}, which offers persona-aligned dialogue samples. It uses a BERT-based\footnote{\url{https://huggingface.co/bert-base-uncased}} pipeline to extract task-relevant entities and constructs structured topic–relation–object triples through topic modeling, guided by a predefined taxonomy of tasks and topics. This extractor is lightweight and deployable in CPU-based environments with modest memory usage.

The Community-based Recommender retrieves related topics and objects from a backend database using the current task context. These are passed back to the PersoAgent, which aligns the results with the latest dialogue topic to generate context-aware responses. Although the PersoAgent executes via the OpenAI API and is not locally deployed, this setup enables consistent tool invocation and reliable formatting, capabilities that are often unstable in similarly sized open-source models during multi-step reasoning.

A structured prompt is engineered to guide the PersoAgent's behavior during interaction. The prompt begins by briefly introducing the task context, defining the agent’s role, and presenting relevant user information along with the current task at hand. It then specifies the available tools, describing their purpose and the conditions under which each should be used. Finally, a few-shot learning strategy is applied, providing one illustrative example per tool to demonstrate proper usage. The prompt concludes with explicit instructions for the agent to strictly follow a predefined output structure (a JSON object), ensuring seamless integration with the FastAPI application backend.

\begin{figure*}[!h]
\centering
  \includegraphics[width=0.64\textwidth]{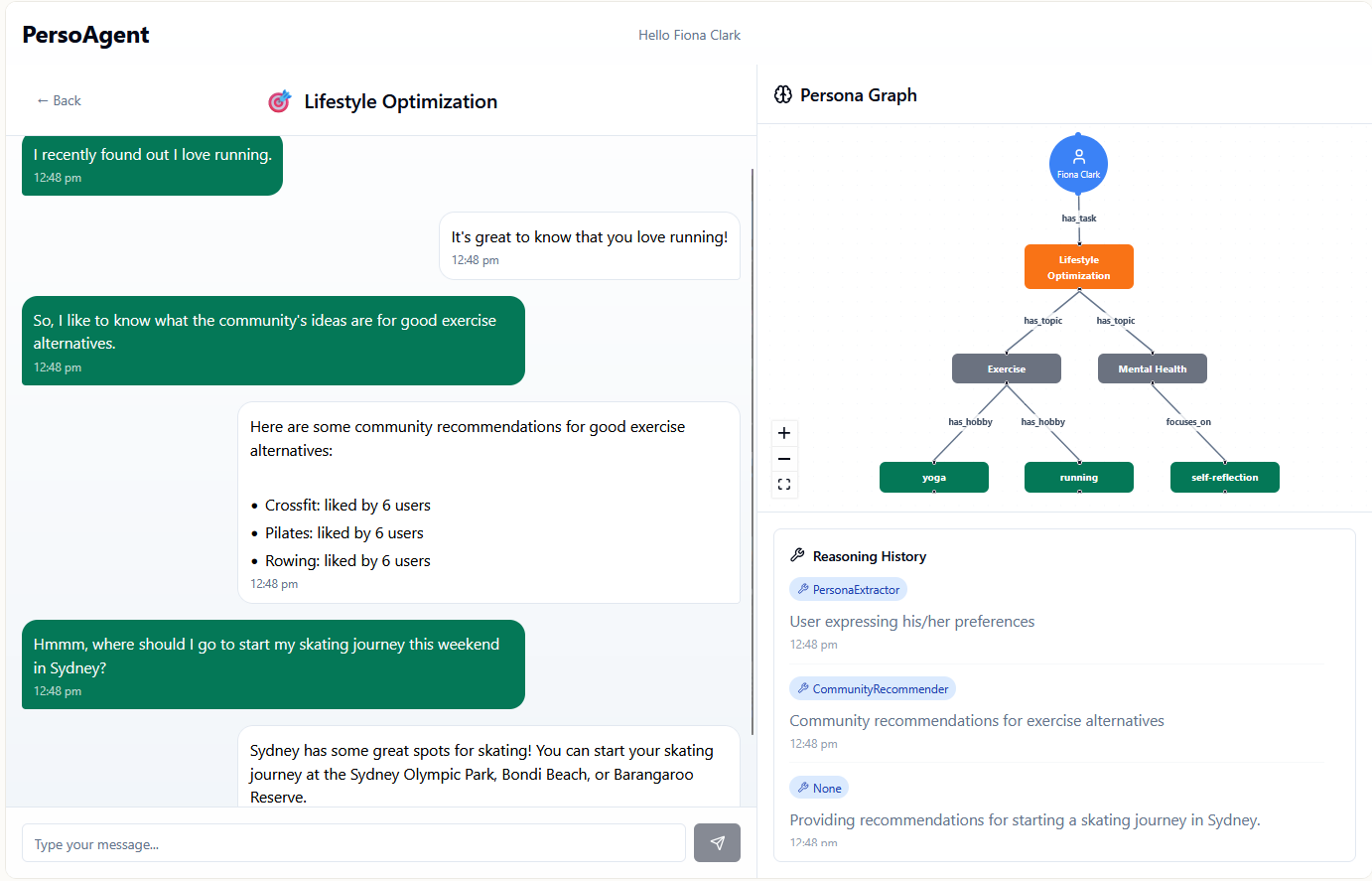}
  \caption{User interface for the help seeker showing personalized assistance features.}
  \label{fig:general-user-interface}
\end{figure*}

\textbf{Analytic Tools.} This module includes a \textit{Labeling Assistant} and a similarity-based \textit{Classifier}. Persona summaries and classification tasks are passed to Phi-4-mini-instruct\footnote{\url{https://huggingface.co/microsoft/Phi-4-mini-instruct}}, a lightweight instruction-tuned LLM developed by Microsoft. With under 4 billion parameters, Phi-4 mini supports efficient few-shot reasoning and chain-of-thought (CoT) explanation generation. It is guided by structured prompts defining task logic and format, and runs smoothly on a single 8GB GPU, making it ideal for analyst-facing deployment without significant resource demands.

In parallel, a TF-IDF–based classifier computes cosine similarity between persona texts and predefined category vectors, offering a fast, interpretable cross-check of LLM-generated labels. Together, the analytic pipeline, consisting of Phi-4 mini, the BERT extractor, and the TF-IDF classifier, supports explainable annotation and iterative refinement.

The system is deployed as a modular application with two frontends: a FastAPI\footnote{\url{https://fastapi.tiangolo.com/}} backend for programmatic access to the core components, and a React-based\footnote{\url{https://reactjs.org/}} web interface for end-user interaction and analyst supervision. This dual implementation facilitates both rapid prototyping and user-centric evaluation of system functionality.

\subsection{Preliminary Evaluation}
We conducted a preliminary evaluation using the UniEval framework to assess the quality of system-generated responses, focusing on key response-level aspects. Specifically, we evaluated Naturalness, Coherence, Groundedness, and Understandability across a limited set of prompts and outputs. The framework achieved average scores of 0.84 for Naturalness, 1.0 for Coherence, 1.0 for Groundedness, and 0.86 for Understandability. These results indicate the responses are fluent, well-structured, and closely aligned with the user queries.\footnote{\href{https://github.com/salehafzoon/persopilot-backend/blob/main/Evaluation/evaluation.ipynb}{github.com/salehafzoon/persopilot-backend/Evaluation/evaluation.ipynb}}

\section{Demo Scenarios} 

\textbf{Scenario 1: Task-Driven Support for Help Seekers.} 

In this scenario, a help seeker engages with the PersoAgent to receive personalized assistance aligned with a specific task. Upon login, the user selects from predefined personalization-driven tasks, such as \textit{Content Consumption}, \textit{Lifestyle Optimization}, or \textit{Career Development}, each accompanied by a brief description and a list of relevant topics.

After choosing a task (e.g., \textit{Lifestyle Optimization}), the user is directed to the main interface shown in Fig.~\ref{fig:general-user-interface}, which consists of a natural language chat panel (left), a task-filtered persona graph (top-right), and a reasoning panel (bottom-right). The persona graph presents a dynamic subgraph of user traits filtered by the selected task, while the reasoning panel explains the agent’s rationale and tools invoked per response.
When the user expresses a preference (first message in Fig.~\ref{fig:general-user-interface}), PersoAgent applies its persona extraction module to generate a contextualized entry, which is then added to the persona graph in real time. This allows the user to inspect how their preferences are interpreted. Simultaneously, the agent delivers a personalized reply in the chat panel, and the reasoning panel logs the tool used along with its justification.
If the user requests community insights (second message), PersoAgent activates its community-based recommender to retrieve aggregated preferences from similar users within the same task. These recommendations are filtered by the ongoing conversation topic and displayed directly in the chat panel, with the reasoning panel documenting the filtered topic and recommendation strategy.
For general information-seeking queries (third message), where tool invocation is unnecessary, PersoAgent generates a direct response grounded in contextual understanding. The reasoning panel reflects that the request was handled without external tools, ensuring transparency even in non-personalized interactions.

\textbf{Scenario 2: Analyst-Guided Persona Classification Workflow}

\begin{figure*}[!h]
    \centering
    \begin{subfigure}[b]{0.47\textwidth}
        \centering
        \includegraphics[width=\textwidth]{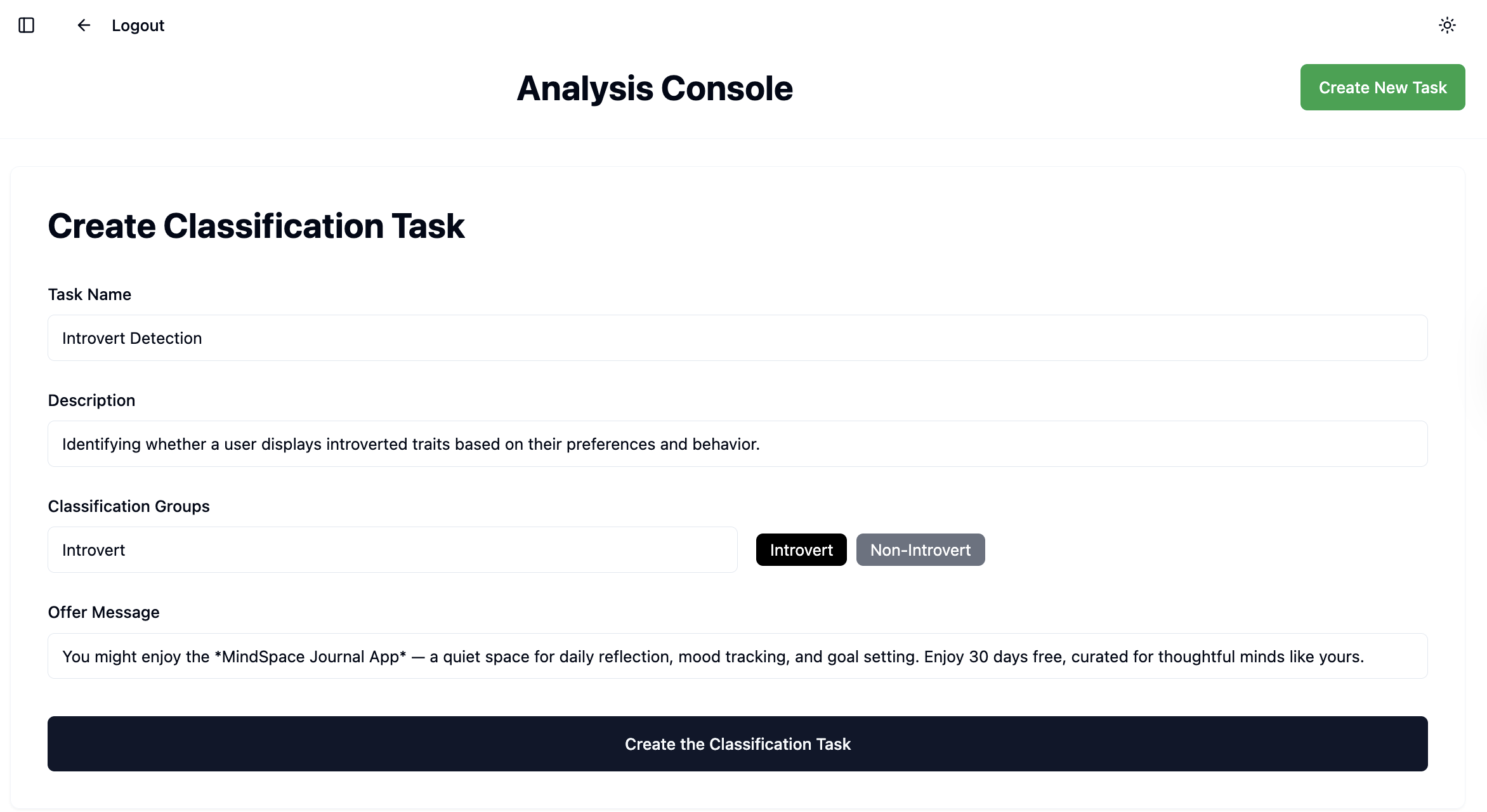}
        \caption{Classification task creation interface.}
        \label{fig:subfig1}
    \end{subfigure}
    \hfill
    \begin{subfigure}[b]{0.47\textwidth}
        \centering
        \includegraphics[width=\textwidth]{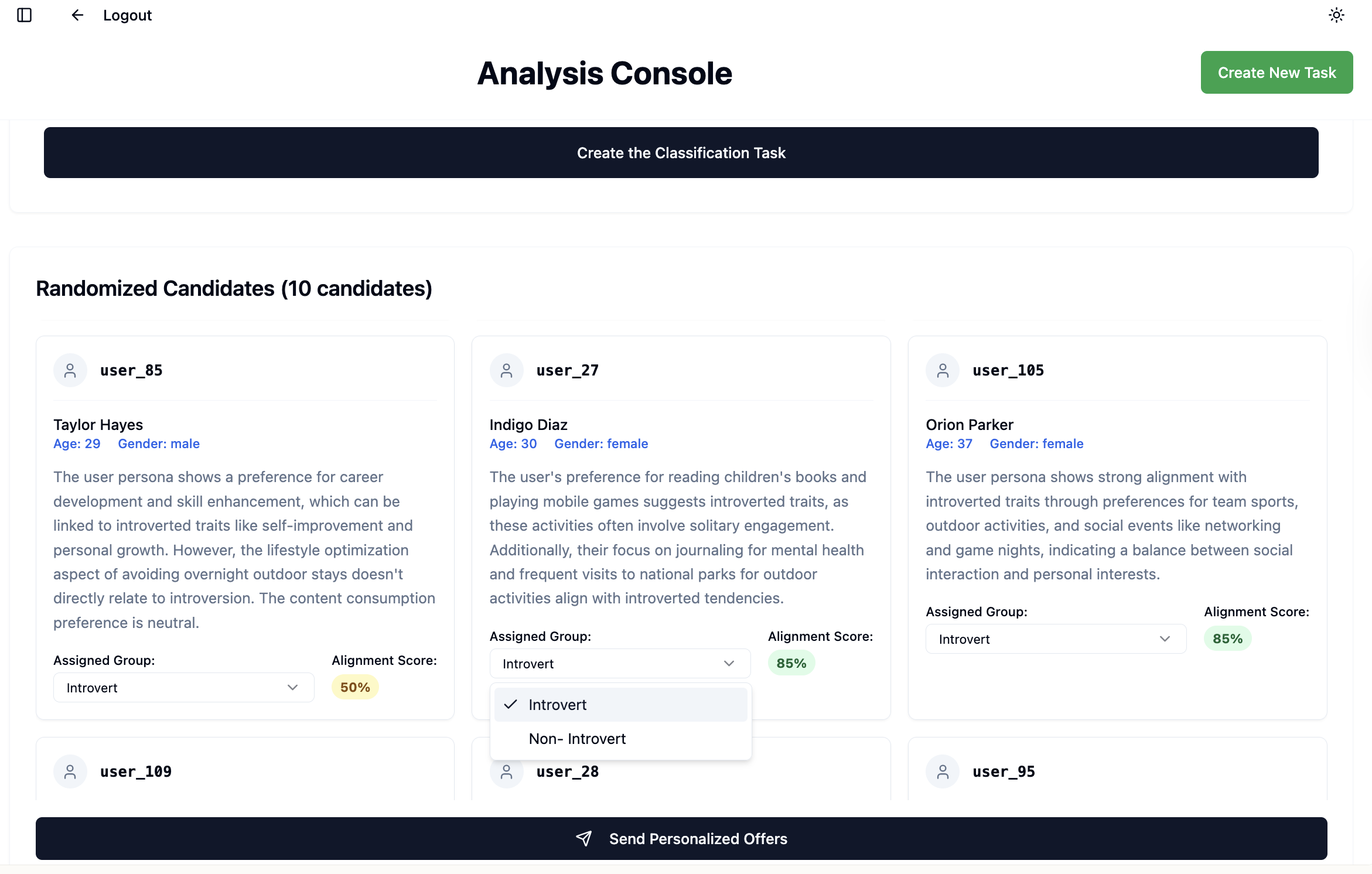}
        \caption{Labeling assistant response with justification.}
        \label{fig:subfig2}
    \end{subfigure}

    \caption{Analyst workflow showing (a) classification task setup, and (b) labeling assistant output for analyst review, update, and confirmation.}
    \label{fig:clf_creation_and_labelling}
\end{figure*}

Figure~\ref{fig:clf_creation_and_labelling} illustrates the analyst's interaction with the Analytic Tools. The workflow begins with the analyst defining a new binary classification task (e.g., \textit{Introvert Detection}) along with a personalized offer message targeting users classified under the positive label (Fig.~\ref{fig:clf_creation_and_labelling}a). Once the task is created, the labeling assistant iterates over system-generated persona summaries, structured by task and topic, and proposes a predicted label for each user, along with a confidence score and justification (Fig.~\ref{fig:clf_creation_and_labelling}b). 
For instance, a candidate may be shown with basic demographic attributes, a confidence score of 0.85, and a justification aligned with the classification criteria. By default, users with a score $\geq 0.60$ are assigned to the positive group, while others are assigned to the negative class. All suggestions appear in an analyst-friendly interface, where labels can be manually reviewed and overridden via a dropdown selector before confirmation.

Once the analyst finalizes the labels, the corresponding personalized offers are automatically sent to the selected users. These offers are displayed on the user’s home page and may be accepted or rejected. Each user response becomes a labeled data point, forming a stream of ground truth feedback that is used to update classifier performance.

\begin{figure}[!h]
\centering
  \includegraphics[width=0.55\textwidth]{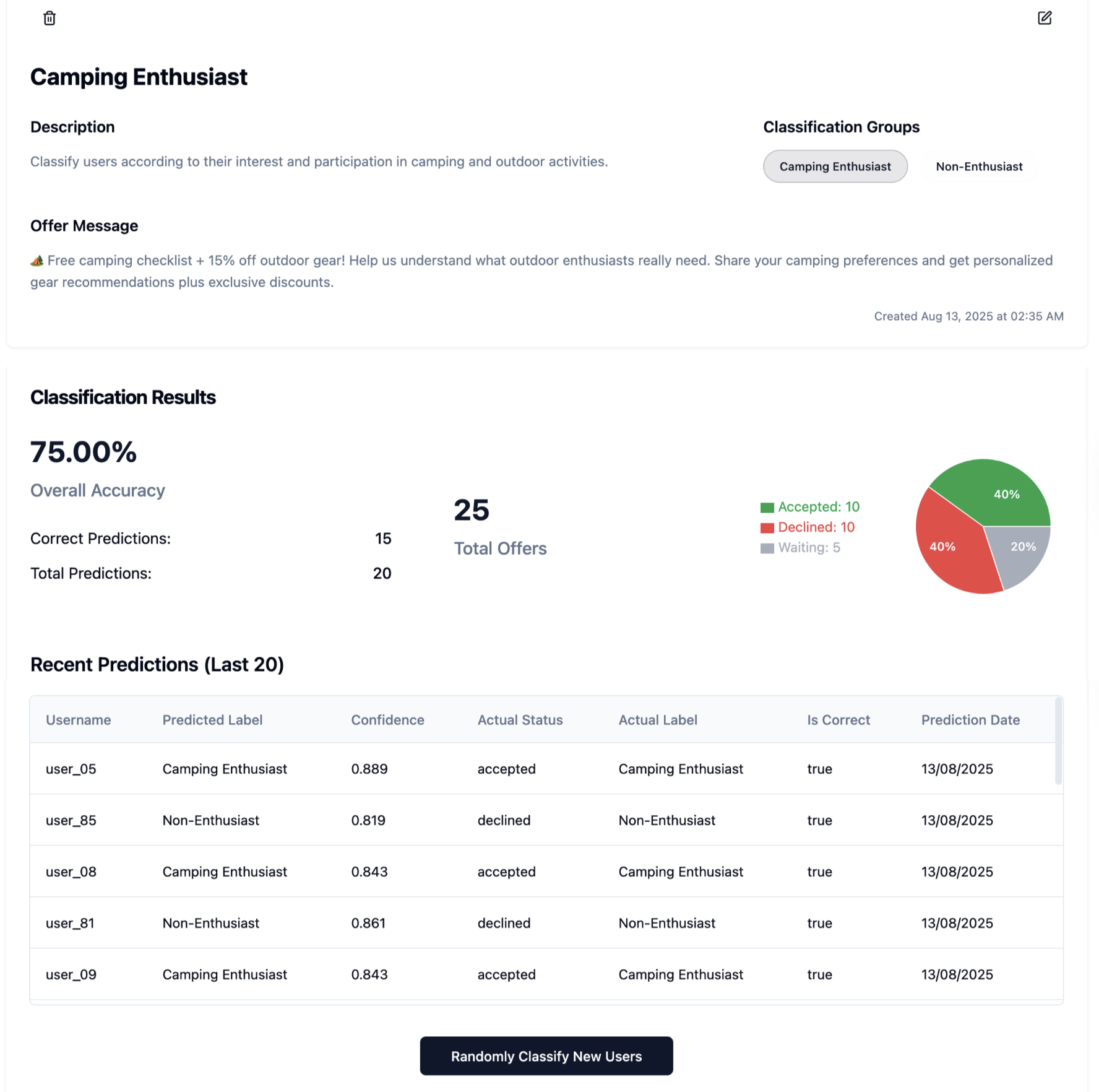}
  \caption{User classification results interface with prediction accuracy, and recent classification outcomes.}
  \label{fig:classification_results}
\end{figure}

As new labels accumulate, the analyst can monitor the evolving statistics via the classification dashboard (Fig.~\ref{fig:classification_results}). This includes real-time updates on user responses—accepted, rejected, or pending. Once sufficient labeled data is available, an additional module in the analysis console is unlocked, enabling automatic classification of new users with a single action (\textit{Randomly Classify New User} button). This module uses a TF-IDF–based similarity approach: contextual persona summaries from labeled users are vectorized and compared against those of unlabeled users to assign classes.

Predictions from this stage trigger the next wave of personalized offers, whose outcomes further refine the model and update the recent prediction tables of the UI. This establishes an active learning loop in which analyst-driven labeling, user interactions, and model updates form a continuous improvement cycle for scalable and adaptive persona classification.

\section*{Acknowledgment}
We acknowledge the Centre for Applied Artificial Intelligence at Macquarie University, Sydney, Australia, for supporting this research. We also acknowledge the use of AI-assisted language tools for editorial refinement during the preparation of this manuscript.

\bibliographystyle{unsrt} 
\bibliography{references}

\end{document}